\documentclass[useAMS,usenatbib]{pasj02}
\usepackage{tikz}
\usepackage{lscape}
\usepackage{color}
\usepackage{comment}
\usepackage{graphicx}
\usepackage{natbib, aas_macros}
\usepackage{bm}
\usepackage{placeins}
\usepackage{amsmath}
\usepackage{cuted}
\usepackage{flushend}

\usepackage{soul}
\sethlcolor{pink}
\usepackage[switch,mathlines]{lineno} 

\newcommand{\simgt}{\lower.5ex\hbox{$\; \buildrel > \over \sim \;$}}
\newcommand{\simlt}{\lower.5ex\hbox{$\; \buildrel < \over \sim \;$}}

\def\btheta{\mbox{\boldmath $\theta$}}

\def\h70kpc{\mathrel{h_{70}^{-1}{\rm kpc}}}

\def\h70Msol{\mathrel{h_{70}^{-1}M_\odot}}
\usepackage{version}

\begin{document}

\Received{}%{yyyy/mm/dd}
\Accepted{}%{yyyy/mm/dd}
%\Published{yyyy/mm/dd}

%\author{et al.}

%\altaffiltext{3}{C-Address of Institute}
%\email{ccccc@xxx.xxx.xx.xx}
%%% end:list of authors

\Received{}%{yyyy/mm/dd}
\Accepted{}%{yyyy/mm/dd}
%\Published{yyyy/mm/dd}

\Received{}%{yyyy/mm/dd}
\Accepted{}%{yyyy/mm/dd}
%\Published{yyyy/mm/dd}

\title{XRISM–Subaru views of Abell 754: an off-axis, near-line-of-sight merging cluster\thanks{ Based on data collected at Subaru Telescope, which is operated by the National Astronomical Observatory of Japan.}}

%%% begin:list of authors
% Do NOT capitalize all letters in "textsc".
\author{Nobuhiro \textsc{Okabe}\altaffilmark{1,2,3}\orcid{0000-0003-2898-0728}}
\email{okabe@hiroshima-u.ac.jp}
\altaffiltext{1}{Physics Program, Graduate School of Advanced Science and Engineering, Hiroshima University, 1-3-1 Kagamiyama, Higashi-Hiroshima, Hiroshima 739-8526, Japan}
\altaffiltext{2}{Hiroshima Astrophysical Science Center, Hiroshima University, 1-3-1 Kagamiyama, Higashi-Hiroshima, Hiroshima 739-8526, Japan}
\altaffiltext{3}{Core Research for Energetic Universe, Hiroshima University, 1-3-1, Kagamiyama, Higashi-Hiroshima, Hiroshima 739-8526, Japan}

\author{Yuki \textsc{Omiya}\altaffilmark{4} \orcid{0009-0009-9196-4174}}
\altaffiltext{4}{Department of Physics, Nagoya University, Aichi 464-8602, Japan}

\author{Kazuhiro \textsc{Nakazawa}\altaffilmark{4} \orcid{0000-0003-2930-350X}}

\author{Naomi \textsc{Ota}\altaffilmark{5} \orcid{0000-0002-2784-3652}}
\altaffiltext{5}{Department of Physics, Nara Women's University, Nara 630-8506, Japan}

\author{Nhan T. \textsc{Nguyen-Dang}\altaffilmark{5} \orcid{0000-0001-6178-7714}}

\author{Yuto \textsc{Ichinohe}\altaffilmark{6} \orcid{0000-0002-6102-1441}}
\altaffiltext{6}{RIKEN Nishina Center, Saitama 351-0198, Japan}

\author{Shutaro \textsc{Ueda}\altaffilmark{7, 8} \orcid{0000-0001-6252-7922}}
\altaffiltext{7}{Faculty of Mathematics and Physics, Institute of Science and Engineering, Kanazawa University, Kakuma, Kanazawa, Ishikawa, 920-1192 Japan}
\altaffiltext{8}{Advanced Research Center for Space Science and Technology (ARC-SAT), Kanazawa University, Kakuma, Kanazawa, Ishikawa, 920-1192, Japan}

%% `\KeyWords{}' always has to be placed before `\maketitle'.
 %Do NOT move this preamble from here!

\maketitle

\begin{abstract}
 We report a weak-lensing (WL) mass measurement for the merging cluster Abell 754 and impose constraints on the merger trajectory. The trajectory analysis adopts a two-body model with a point-mass approximation and dynamical friction, refined using numerical simulations of major mergers and characterized by Euler angles. We first conduct WL analysis using the two-dimensional (2D) shear pattern from the Subaru Hyper Suprime-Cam (HSC) in combination with Suprime-Cam images to assist in color selection. 
 The WL mass map shows a distinct double-peak structure located around the western and eastern brightest cluster galaxies (BCGs), as reported in the literature. 
The two-halo component analysis, which utilizes the 2D shear pattern over the cluster entire region and considers the lensing covariance matrix from uncorrelated large-scale structures, indicates mass values of $M_{200}^W=3.13_{-1.00}^{+1.53}\times10^{14}h_{70}^{-1}M_\odot$ and $M_{200}^E=6.41_{-1.97}^{+2.92}\times10^{14}h_{70}^{-1}M_\odot$. Thus, the eastern mass component associated with the X-ray tadpole-shaped gas is the main cluster. No substantial structural components are detected in the line-of-sight velocities of the member galaxies. 
By combining the WL parameters, line-of-sight velocities, X-ray morphology, and priors informed by X-ray kinematics, we estimate an impact parameter of approximately 0.77 Mpc at an initial separation of 2 Mpc from the main cluster. The merger plane is inclined at about 20 degrees relative to the line-of-sight. Interestingly, this system is an off-axis, near-line-of-sight merger. This characteristic arises because the trajectory within the merger plane is altered during the pericenter passage, causing the apparent motion to transition from predominantly along the line-of-sight before the core passage to mainly within the plane of the sky afterward.
This study will assist in conducting numerical simulations to understand the {\it XRISM} observations.
\end{abstract}

%\pagewiselinenumbers
\clearpage %

%\linenumbers

\section{Introduction}

Galaxy clusters, encompassing dark matter, hot gas, and galaxies, are the universe's most massive gravitationally bound structures. During cluster mergers, which release gravitational energy on the scale of $10^{64}\,{\rm erg}$, the gas experiences significant and rapid transformations. These changes are driven by the gas's collisional nature, leading to hydrodynamic compression, instabilities, and shocks. Consequently, this results in a shorter lifetime of the gas halo than that of the dark matter halo and non-linear variations in X-ray measurements.
Although cluster mergers are intermittent events, such a non-linear increase, the so-called ``merger boosts'', affects both the thermal history and the cluster evolution \citep[e.g.][]{2001ApJ...561..621R}.

An observational understanding of cluster mergers requires complementary multi-wavelength data of X-ray, Sunyaev-Zel'dovich, weak gravitational lensing (WL) analysis, optical, and radio \citep[e.g.][]{2019PASJ...71...79O,2021MNRAS.501.1701O} and their joint analysis to resolve their spatial structures sufficiently. 
One of the important observational techniques is weak gravitational lensing (WL) analysis, because the WL analysis enables us to map out projected mass distributions without any assumptions of dynamical states and mass models. Dark matter halos, which are a dominant mass component, survive longer than the gas halos during cluster mergers due to their collisionless nature. Therefore, even though X-ray surface brightness distributions are highly disturbed, the WL mass map can identify two mass components of merging clusters \citep[e.g.][]{2008PASJ...60..345O}. Furthermore, WL mass distribution enables us to deduce merger dynamics under the assumption of a point mass approximation, which is complementary to measuring the line-of-sight gas motion by superb high spectroscopic resolution of \texttt{Resolve} on board of {\it XRISM}.

This paper reports the WL mass measurement for a well-known merging cluster, Abell 754 \citep[e.g.][]{1996ApJ...466L..79H,2003ApJ...586L..19M,2003AstL...29..425K,2004ApJ...615..181H,2008PASJ...60..345O,2011ApJ...728...82M,2024A&A...690A.222B}, at the redshift $z=0.052921$. There are two brightest cluster galaxies (BCGs); WISEA~J090919.21$-$094159.0  ($z_{\rm EB} = 0.054291$, hereafter E-BCG) in the eastern region associated with the tadpole-shaped gas and WISEA~J090832.37$-$093747.3 ($z_{\rm WB} = 0.054841$, hereafter W-BCG) in the western region (Figure \ref{fig:massmap}). The optical luminosity of the W-BCG is brighter than that of the E-BCG (Figure \ref{fig:massmap} and \citet{2008PASJ...60..345O})

\citet{2008PASJ...60..345O} carried out WL analysis on the central region ($34'\times28'$) using Subaru/Suprime-Cam and found a double peak structure around W-BCG and E-BCG. The highest peak in the WL mass map corresponds to the western mass clump linked to W-BCG.
However, they could not measure the masses for the two components because the FoV is limited and the methodology had not yet been developed. Based on the visual interpretation of the WL mass map and the {\it Chandra} X-ray image, they speculated that the western mass component represents the primary cluster.

In this study, we perform a two-dimensional (2D) WL analysis focusing on multiple halo structures, as outlined by \citet{2011ApJ...741..116O}. This investigation leverages the 2D shear pattern methodology \citep{Oguri10b}, an approach whose techniques were developed after \citet{2008PASJ...60..345O}. Our analysis incorporates observational data retrieved from the Subaru Telescope with the Hyper Suprime-Cam (HSC) \citep[HSC;][]{2018PASJ...70S...1M}, which provides extensive coverage of the observational field. Furthermore, we augment our dataset by integrating it with pre-existing Suprime-Cam observations, facilitating a secure selection of background galaxies essential for our study.

This paper is organised as follows: Secs. \ref{sec:data} and \ref{sec:result} describe the data analysis and present the results and discussion, respectively.  Sec \ref{sec:summary} is devoted to the summary. We assume a flat $\Lambda$CDM cosmology with the cosmological parameters of $H_0=70\,{\rm km\,s^{-1}Mpc^{-1}}$, $\Omega_{m,0}=0.3$, and $\Omega_{\Lambda,0}=0.7$.

\section{Data analysis} \label{sec:data}

\subsection{Data reduction and shape measurement}
We retrieved the archival HSC $i2$'-band image covering $2.5$ deg$^2$. The data were reduced by \texttt{hscpipe} version 8 \citep{2018PASJ...70S...5B} using the standard process of single-visit processing, mosaicking, and coadding images with photometric calibration of the Pan-STARRS1 Survey \citep{2012ApJ...756..158S}. 
The brighter-flatter effect is corrected by flat-fielding. The exposure time of the reduced image is $50$ minutes, and the seeing size is $\sim 0.83''$.  

For a background selection, we also used $g$ and $R_{\rm c}$ Suprime-Cam imaging \citep{2008PASJ...60..345O}, covering the central region of $0.3\,{\rm deg}^2$ in the HSC area, corresponding to $1.1$ Mpc radius.
Because the dilution effect predominantly occurs in the central area, where the number fraction of centrally concentrated member galaxies relative to uniformly distributed background galaxies rises with decreasing radius \citep[e.g.][]{2005ApJ...619L.143B}, the Surprime-Cam imaging enables us to effectively reduce the contamination of member galaxies in the source catalog. We measured galaxy shapes based on a modified version \citep{2016MNRAS.461.3794O} of the KSB method \citep{1995ApJ...449..460K}. Because the HSC imaging covers a wide field, the half-light radii of stellar sources vary with their positions. Shape measurements are performed on patch images, each with dimensions of 4200 pix $\times$ 4200 pix, which are generated by \texttt{coaddDriver.py} in \texttt{hscpipe}, and the resulting measurements are subsequently merged into a unified shape catalog. The same galaxies that appear in adjacent patch images are utilized to check the accuracy of shape measurements, and their average values are recorded in the shape catalog. 

We measured the image ellipticity, $e_\alpha({\bm \theta})$, from the weighted quadrupole moments of the surface brightness of objects at the position of ${\bm \theta}$. The PSF anisotropy is corrected by solving $e_\alpha'({\bm \theta})=e_\alpha({\bm \theta})-P_{\rm sm}^{*\alpha\beta}q^*_\beta({\bm \theta})$. In this expression, $q^*_\alpha({\bm \theta})$ is given by $(P_{\rm sm})^{-1}_{\alpha\beta} e^{*\beta}$, where $P_{\rm sm}^{\alpha\beta}$ represents the smear polarizability tensor, and all quantities marked with an asterisk refer to stellar sources. At each galaxy location, ${\bm \theta}$, we determined $q^*_\alpha({\bm \theta})$ by applying a second-order bi-polynomial fitting function within each patch image, combined with iterative $\sigma$-clipping.
The anisotropic PSF correction is assessed by residual stellar ellipticity, its correlation function, and the cross-correlation function between corrected galaxy and residual stellar ellipticities. We do not find a significant positive or negative correlation in each patch. The average of residual stellar ellipticity becomes $\langle \delta e_1^* \rangle=(0.001\pm1.1)\times 10^{-3}$
and $\langle \delta e_2^* \rangle=(-0.007\pm1.1)\times 10^{-3}$ from the raw ellipticity $\langle e_1^* \rangle=-0.003\pm0.010$ and $\langle e_2^* \rangle=0.011\pm0.007$.

We then corrected the isotropic smearing effect of galaxy shapes due to seeing and the Gaussian window function used for the shape measurements. As reported in previous studies \citep[e.g.][]{2014ApJ...795..163U,2015MNRAS.449..685H,2016MNRAS.461.3794O}, the pre-seeing shear polarizability tensor is very noisy for individual faint galaxies because of its non-linearity.  The calibration of the shear polarizability tensor depending on the Gaussian smoothing radius, $r_g$, employs a similar technique as outlined by \citet{2016MNRAS.461.3794O}. We apply the scalar correction approximation to the shear polarizability tensor. We then select galaxies with a high signal-to-noise ratio of $\nu>50$ and calibrate the $r_g$ dependence on the polarizability tensor through a ground-based image simulation employing Shear Reconvolution Analysis \citep[\texttt{Shera};][]{2012MNRAS.420.1518M}. 
We then interpolate the polarizability tensor for individual galaxies with $\nu>10$ using $r_g$ and the absolute value of the ellipticity, $|e|$.

The photometric redshift of each galaxy was estimated from an average of 50 neighboring COSMOS2025 galaxies \citep{2025arXiv250603243S} in the $i2'$ or $i2'g'R_{\rm c}$ magnitude. The uncertainty of the photometric redshift is the standard error of the mean. 
We excluded galaxies located within the red-sequence in the $g'-R_{\rm c}$ versus $R_{\rm c}$ plane. We note that faint and small galaxies in the source catalog are predominantly background galaxies, since Abell 754 is located at a very low redshift. After this procedure, the number densities in the annuli of $0\!-\!0.5\,h_{70}^{-1}\,{\rm Mpc}$ and $0.5\!-\!1\,h_{70}^{-1}\,{\rm Mpc}$ centered on the E-BCG are reduced by approximately $4\%$ and $2\%$, respectively. The number-density ratio inside versus outside a radius of $0.9\,h_{70}^{-1}\,{\rm Mpc}$ from the Suprime-Cam field center is $0.98$, demonstrating that there is no statistically meaningful enhancement toward the center. 

The resulting number density of the background galaxies is $n_g\simeq19\,{\rm arcmin}^{-2}$.

\subsection{Weak-mass reconstruction}

We computed the dimensional surface mass density, $\Sigma(\boldsymbol{\theta})$, using a Fourier transform with zero-padding, applying the Kaiser \& Squires (KS) inversion method \citep{1993ApJ...404..441K}. In this context, we consider the weak limit of the reduced dimensional shear, represented as $\Delta \Sigma = \Delta \tilde{\Sigma}/(1-{\mathcal L}\Sigma)\simeq \Delta \tilde{\Sigma}$, where $\mathcal{L}$ corresponds to the inverse of the critical surface mass density,  $\Sigma_{{\rm cr}}= c^2D_{s}/4\pi G D_{l} D_{ls}$. Here, $D_s$ and $D_{ls}$ are the
angular diameter distances from the observer to the sources and from the
lens to the sources, respectively.

We pixelized the shear distortion data in a regular grid of pixels using a Gaussian kernel, $G(\btheta)\propto \exp[-\btheta^2/(2\sigma_g^2)]$ with ${\rm FWHM}=2\sqrt{2\ln 2}\sigma_g$. 
The dimensional shear field, $\Delta \tilde{\Sigma}_{\alpha}(\btheta)$, at angular position $\btheta$ was obtained by 
\begin{eqnarray}
\Delta \tilde{\Sigma}_{\alpha}(\btheta) = 
\frac{\sum_{i} e_{\alpha ,i}G(\btheta_i-\btheta) w_{i}  \Sigma_{{\rm cr}}(z_{l}, z_{s,i})}{\sum_{i}  G(\btheta_i-\btheta) w_{i}}, \label{eq:massmap}
\end{eqnarray}
where $\alpha=1,2$ of the two component of the  ellipticity, $i$ denotes the $i$-th galaxy, and $w_i=1/({e_{{\rm err},i}^2+\alpha^2})/\Sigma_{{\rm cr}}(z_{l}, z_{s,i})^2$ is the weighting function with a softning constant $\alpha=0.4$. 
The noise map, $\sigma_{\Sigma}$, was estimated through $N_{\rm boot}=5000$ bootstrap iterations by rotating the ellipticity of galaxies and varying source redshifts within their $1\sigma$ uncertainties while maintaining fixed positions.
The resulting signal-to-noise ratio map is $S/N=\Sigma/\sigma_\Sigma$. We use FWHM=3 arcmin.

\subsection{2D WL analysis}

To measure WL masses of the main and subclusters, we carried out two-dimensional shear analysis \citep{Oguri10b} using multi mass components \citep[e.g.][]{2011ApJ...741..116O,2025A&A...700A..46O}. We assume a spherical NFW profile \citep{NFW96} specified by 
\begin{equation}
\rho_{\rm NFW}(r)=\frac{\rho_s}{(r/r_s)(1+r/r_s)^2},\label{eq:rho_nfw}
\end{equation}
where $\rho_s$ is the central density parameter and $r_s$ is the scale
radius. We used two parameters of the spherical mass, $M_\Delta$, and the halo concentration, $c_\Delta$, instead of $\rho_s$ and $r_s$.
The spherical mass and the halo concentration are defined by
\begin{eqnarray}
    M_{\Delta}=\frac{4}{3}\pi \Delta \rho_{\rm cr} r_{\Delta}^3,  \quad\quad    c_{\Delta}=\frac{r_{\Delta}}{r_s},  
\end{eqnarray}
respectively. Here, $\rho_{\rm cr}(z)$ is the critical mass density at the cluster redshift and $\Delta$ is the overdensity.  

The mean dimensional shear, denoted as $\Delta \Sigma_{\alpha}$, for each grid cell is calculated in a manner akin to equation \ref{eq:massmap}. However, in this context, $G$ and $\Delta \tilde{\Sigma}$ is substituted with $\mathcal{H}(\btheta-\btheta_k)-\mathcal{H}(\btheta - \btheta_{k+1})$ and $\Delta \Sigma$, respectively. Here, $\mathcal{H}$ is the Heaviside step function and $\btheta_k$ represents the $k$-th angular coordinate. The representative position in each grid cell was estimated by a weighted harmonic mean \citep{2016MNRAS.461.3794O}. The size of each grid cell is $2$ arcmin. The inversion of critical surface mass density is given by
\begin{eqnarray}
    \mathcal{L}(\btheta)=\frac{\sum_i \Sigma_{\rm cr}(z_l,z_{s,i})^{-1}w_i}{\sum_i w_i}.
\end{eqnarray}

The likelihood was calculated by following \citet{Oguri10b}. We concatenated the vector with $\alpha=1$ and one with $\alpha=2$ vertically into a single vector, as follows, 
\begin{eqnarray}
    {\bm v}&=&\begin{pmatrix}
        \Delta \Sigma_1(\btheta_k) - g_1^{\rm model}(\btheta_k) \\
        \Delta \Sigma_2(\btheta_k) - g_2^{\rm model}(\btheta_k) \\
          \end{pmatrix}, 
    \end{eqnarray}
where $g_\alpha^{\rm model}$ is the dimensional shear pattern of the NFW model at the $k$-th grid cell. 
The error covariance matrix is composed of the shape noise, the uncertainty of photometric redshift, and the uncorrelated large-scale structure lensing covariance matrix, as follows,
\begin{eqnarray}
    {\bm C}={\bm C}^{\rm shape}+{\bm C}^{\rm s}+{\bm C}^{\rm LSS}, \label{eq:covariance}
\end{eqnarray}
where each $\bm C$ is specified as
\begin{equation}
    {\bm C}= \begin{pmatrix}
{\bm C}_{11} & {\bm C}_{12} \\
{\bm C}_{12} & {\bm C}_{22} \\
\end{pmatrix}.
\end{equation}
Here, ${\bm C}^{\rm shape}_{\alpha\beta}$ is a block matrix of the ellipticity component of $\alpha$ and $\beta$. 
The shape noise at positions $\btheta_k$ and $\btheta_m$ is described by ${\bm C}^{\rm shape}_{\alpha\beta,km}=\sigma_{\Delta \Sigma}^2\delta_{\alpha\beta}\delta_{km}$. The shape noise is estimated through 5000 bootstrap iterations by rotating the ellipticities.
Similarly, the uncertainty of the photometric redshift is ${\bm C}^{\rm s}_{\alpha\beta,km}=\sigma_{s,\alpha\beta}^2\delta_{\alpha\beta}\delta_{km}$ where $\sigma_{s,\alpha\beta}$ is the photometric uncertainty computed by error propagation. The uncorrelated LSS covariance matrix \citep{Oguri10b, 2001PhR...340..291B} is given by 
\begin{eqnarray}
    C_{11}^{\rm LSS}&=&(\cos 2\phi)^2 \xi_{++}(r)+(\sin 2\phi)^2 \xi_{\times\times}(r) \\ 
    C_{22}^{\rm LSS}&=&(\sin 2\phi)^2 \xi_{++}(r)+(\cos 2\phi)^2 \xi_{\times\times}(r) \\
    C_{12}^{\rm LSS}&=& \sin 2\phi \cos 2\phi (\xi_{++}(r)-\xi_{\times\times}(r)). 
\end{eqnarray}
The shear correlation functions are
\begin{eqnarray}
    \xi_{+}(r) \hspace{-0.5em}&=&\hspace{-0.5em}\frac{\langle \Sigma_{{\rm cr},k}^{-1} \rangle^{-1} \langle \Sigma_{{\rm cr},m}^{-1} \rangle^{-1}}{2\pi} \int^\infty_0 d (\ln \ell)  \ell^2 J_0(\ell r/D_A(z_d)) P_\kappa^{\rm nl}(\ell)  \nonumber \\
    \xi_{-}(r) \hspace{-0.5em}&=&\hspace{-0.5em} \frac{\langle \Sigma_{{\rm cr},k}^{-1} \rangle^{-1} \langle \Sigma_{{\rm cr},m}^{-1} \rangle^{-1}}{2\pi} \int^\infty_0 d (\ln \ell)  \ell^2 J_4(\ell r/D_A(z_d)) P_\kappa^{\rm nl}(\ell) \nonumber \\
    \xi_{\pm}(r)\hspace{-0.5em}&=&\hspace{-0.5em}\xi_{++}(r)\pm \xi_{\times\times}(r) \nonumber
\end{eqnarray}
with a distance of $r=|\btheta_k-\btheta_m|$. We assumed the nonlinear power spectrum using the fitting formula \citep{2003MNRAS.341.1311S} and estimated the lensing power spectral $P_{\kappa}^{\rm nl}$ with a weighting of the source redshift distribution in the entire field. 
The likelihood given the parameters is specified by 
\begin{eqnarray}
    p(\Delta \Sigma_\alpha |{\bm p}) = \prod_{nm} \frac{1}{(2\pi)^{N/2} |{\rm det}{\bm C}|^{1/2}} \exp\left[-\frac{1}{2}{\bm v}_n^T{\bm C}_{nm}^{-1} {\bm v}_m\right],
\end{eqnarray}
where $n$ and $m$ are the indices of the vector ${\bm v}$ and $N$ is the total number of elements. 
Since the covariance matrix is symmetric, the inverse matrix can be described as 
\begin{eqnarray}
 {\bm C}^{-1}&=& \begin{pmatrix}
{\bm T}^{-1} & -{\bm T}^{-1}{\bm C}_{12}{\bm C}_{22}^{-1} \\
-{\bm T}^{-1}{\bm C}_{12}{\bm C}_{22}^{-1} & {\bm C}_{22}^{-1}+{\bm C}_{22}^{-1}{\bm C}_{12}^{\bm T} {\bm T}^{-1}{\bm C}_{12}{\bm C}_{22}^{-1} \\
\end{pmatrix}.    \nonumber \\
{\bm T}&=& {\bm C}_{11}-{\bm C}_{12}{\bm C}_{22}^{-1}{\bm C}_{12}^T \nonumber \\
{\rm det}{\bm C}&=&{\rm det}{\bm C}_{22}{\rm det}{\bm T}. \nonumber 
\end{eqnarray}

We use two halo models of the western and eastern components with $\Delta=200$. Each center is treated as a free parameter. 
We employ the Markov chain Monte Carlo (MCMC) method with the parameter of ${\bm p}=(\ln M_{200}^{E},\ln c_{200}^{E},\alpha^E,\delta^E, \ln M_{200}^{W},\ln c_{200}^{W},\alpha^W,\delta^W)$. Here, $\alpha$ and $\delta$ denote the right ascension and declination of the centroids and the superscripts of $E$ and $W$ denote the eastern and western components, respectively. Since the mass and concentration are positive quantities, the logarithmic quantities for $M_{200}$ and $c_{200}$ are adequate to avoid the artificial boundary at zero. We chose the flat priors on the parameters, as follows; $[\ln 10^{-5}, \ln 50]$ on $\ln M_{200}$, $[\ln 10^{-5}, \ln 30]$ on $\ln c_{200}$, $|\alpha-\alpha_{\rm BCG}|< \Delta $, and $|\delta-\delta_{\rm BCG}|< \Delta$ 
with $\Delta=6$ arcmin.

\section{Result and Discussion} \label{sec:result}

\subsection{Weak-mass reconstruction}

\begin{figure}
    \includegraphics[width=\linewidth]{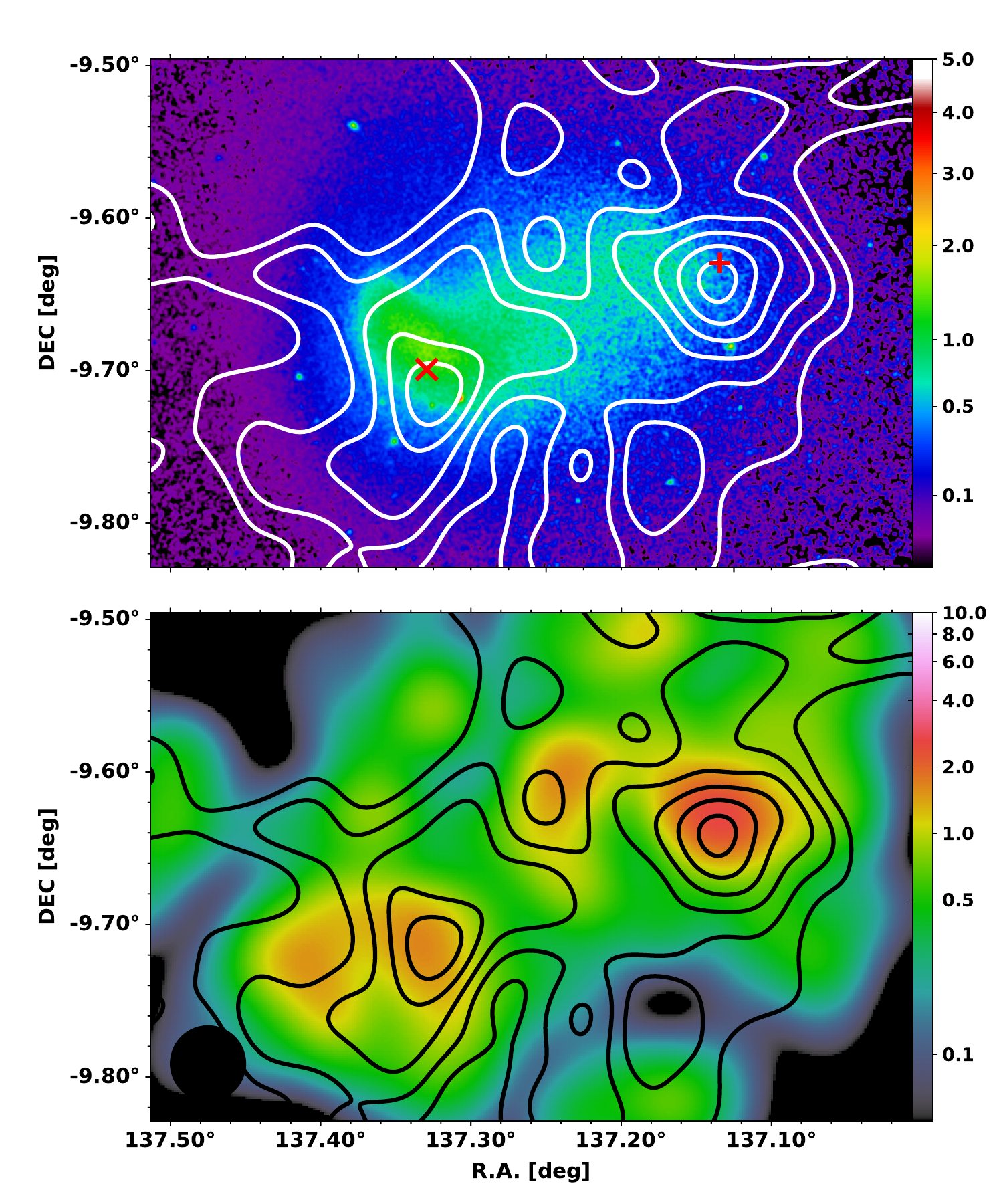}
    \caption{{\it Top}: {\it XMM-Newton} X-ray image ($30'\times20'$) overlaid with the signal-to-noise ratio of the WL mass map, stepped by $1\sigma$ from $1\sigma$. The red $+$ and $\times$ represent the W-BCG ($z_{\rm WB} = 0.054841$) and E-BCG ($z_{\rm EB} = 0.054291$) positions, respectively. A tadpole-shaped gas core is found on the eastern side, with a filamentary gas structure extending northwestward from the core.
    {\it Bottom}: Smoothed $i2'$ luminosity of spec-z member galaxies in a unit of $10^{11}L_\odot$. {Alt text: Two color images aligned vertically.}}
    \label{fig:massmap}
\end{figure}

The contours in Figure \ref{fig:massmap} present the signal-to-noise ratio of the reconstructed mass map, showing a distinct double-peak structure. The most significant peak, corresponding to the western mass component, reaches $6.4\sigma$, while the eastern mass component is linked to the subsequent peak at $5.8\sigma$. This feature is consistent with that found by \citet{2008PASJ...60..345O}. The peak positions coincide with the BCG positions within the smoothing scale. The X-ray tadpole-shaped gas core is associated with the eastern mass component. 
The X-ray filamentary configuration appears to coincidentally connect the mass components to the east and west.

We search for background clusters whose lensing signals are not fully described in the uncorrelated LSS lensing effect, using publicly available cluster catalogs from eRASS1 \citep{2024A&A...685A.106B,2024A&A...688A.210K}, DES-Y3 \citep{2025PhRvD.112h3535A}, and WHL \citep{2012ApJS..199...34W}.
While the cluster region lies beyond the DES-Y3 and WHL areas, it is encompassed by eRASS1; however, no background clusters are detected within the WL mass map region. Therefore, we did not consider background clusters in the following 2D WL analysis.

\subsection{2D WL analysis}

We set the redshift of E-BCG as the main cluster redshift. 
The resulting WL parameters are shown in Table \ref{tab:mass}. The best-fit mass of the eastern cluster is about twice as high as that of the western one, which conflicts with the visual interpretation of the mass map \citep{2008PASJ...60..345O}. It cautions us against relying solely on visual data for interpreting the merger physics. The higher peak of the western mass component, in comparison to its eastern counterpart, can be attributed to its higher concentration parameter superimposed on the main cluster mass distribution. We estimate the mass ratio by randomly picking the cluster masses from the posterior distribution in the following manner, 
\begin{eqnarray}
    f_{\rm ratio}=\exp\left[\frac{1}{N_{\rm sample}}\sum_i(\ln M_{200,i}^W- \ln M_{200,i}^{E} )\right],
\end{eqnarray}
because the logarithmic expression is symmetric under the exchange of $M^W$ and $M^E$; swapping them only changes the sign. Here, $N_{\rm sample}$ is the sampling number. The resulting mass ratio is $f_{\rm ratio}=0.49\pm0.33$. Therefore, Abell 754 is a major merger. The positional separation between the NFW-determined centers is $784\pm73\,h_{70}^{-1}{\rm kpc}$. 

\begin{table}[]
    \centering
    \begin{tabular}{c|c}
 $M_{200}^W$      [$10^{14}h_{70}^{-1}M_\odot$]  & $3.13_{-1.00}^{+1.53}$ \\
  $c_{200}^W$  & $6.41_{-1.88}^{+3.02}$ \\
$\alpha^W$ [degree]& $137.136_{-0.004}^{+0.005}$ \\
 $\delta^W$  [degree]& $-9.638_{-0.005}^{+0.006}$   \\
 $M_{200}^E$    [$10^{14}h_{70}^{-1}M_\odot$]   & $6.41_{-1.97}^{+2.92}$  \\
$c_{200}^E$  & $2.71_{-0.88}^{+1.20}$ \\
$\alpha^E$  [degree]& $137.325_{-0.012}^{+0.011}$ \\
 $\delta^E$[degree] & $-9.712_{-0.026}^{+0.010}$  \\
    \end{tabular}
    \flushleft
    \caption{The results of two-halo components analysis using the 2D shear pattern.}
    \label{tab:mass}
\end{table}

We next computed the total mass, $M_{\Delta}(<r_{\Delta})=M^{\rm E}(<r_{\Delta})+M^{\rm W,off}(<r_{\Delta})$ within the radii, $r_{\Delta}$, from the WL-determined main center, where the mass of the subcluster is estimated by considering an off-centering effect in the three-dimensional space \citep{2021MNRAS.501.1701O} with an assumption of $\sim 0.5\,h_{70}^{-1}{\rm Mpc}$ separation along the line-of-sight (see Sec. \ref{subsec:dis}). The resulting total masses are $M_{200}=10.39_{-2.70}^{+3.24}\times 10^{14}h_{70}^{-1}M_\odot$, $M_{500}= 6.66_{-1.65}^{+1.85}\times 10^{14}h_{70}^{-1}M_\odot$, and $M_{2500}=1.38_{-0.43}^{+0.48}\times 10^{14}h_{70}^{-1}M_\odot$, respectively. 
Under the assumption of zero line-of-sight separation, the inferred total masses become $M_{200}=10.64_{-2.69}^{+3.23}\times 10^{14}h_{70}^{-1}M_\odot$, $M_{500}=7.07_{-1.61}^{+1.81}\times 10^{14}h_{70}^{-1}M_\odot$, and $M_{2500}=1.49_{-0.48}^{+0.53}\times 10^{14}h_{70}^{-1}M_\odot$, respectively.

We repeated WL mass measurements excluding the LSS covariance matrix, and obtained $M_{200}^W=2.80_{-0.66}^{+0.88}\times10^{14}h_{70}^{-1}M_\odot$ and $M_{200}^E=6.44_{-1.35}^{+1.77}\times10^{14}h_{70}^{-1}M_\odot$, giving the mass ratio of $f_{\rm ratio}=0.43\pm0.19$. This indicates that the uncorrelated LSS lensing effect accounts for approximately 50 per cent of the overall error budget. However, it is important to note that this contribution does not lead to a notable alteration in the result.

The uncertainties in the source redshifts within each two-dimensional shear grid contribute approximately $\sim 2\,\%$ to the overall shape noise. Consequently, omitting these redshift uncertainties has only a minor impact on the inferred WL masses, which remain essentially unchanged at $M_{200}^W=3.15_{-0.99}^{+1.57}\times10^{14}h_{70}^{-1}M_\odot$ and $M_{200}^E=6.42_{-1.94}^{+2.95}\times10^{14}h_{70}^{-1}M_\odot$. This is due to the low lensing efficiency caused by the nearby redshift of the galaxy cluster.

Accordingly, in the covariance matrix (eq. \ref{eq:covariance}) for the WL mass estimates, the error contributions from shape noise and from uncorrelated LSS lensing are of comparable magnitude, while the impact of the source redshift term is much smaller. The importance of the uncorrelated LSS lensing effect on WL mass estimates for nearby clusters is also discussed in \citet{2003MNRAS.339.1155H}.

\subsection{Spectroscopic redshifts of members galaxies} \label{subsec:specz}

\begin{figure}
    \includegraphics[width=\linewidth]{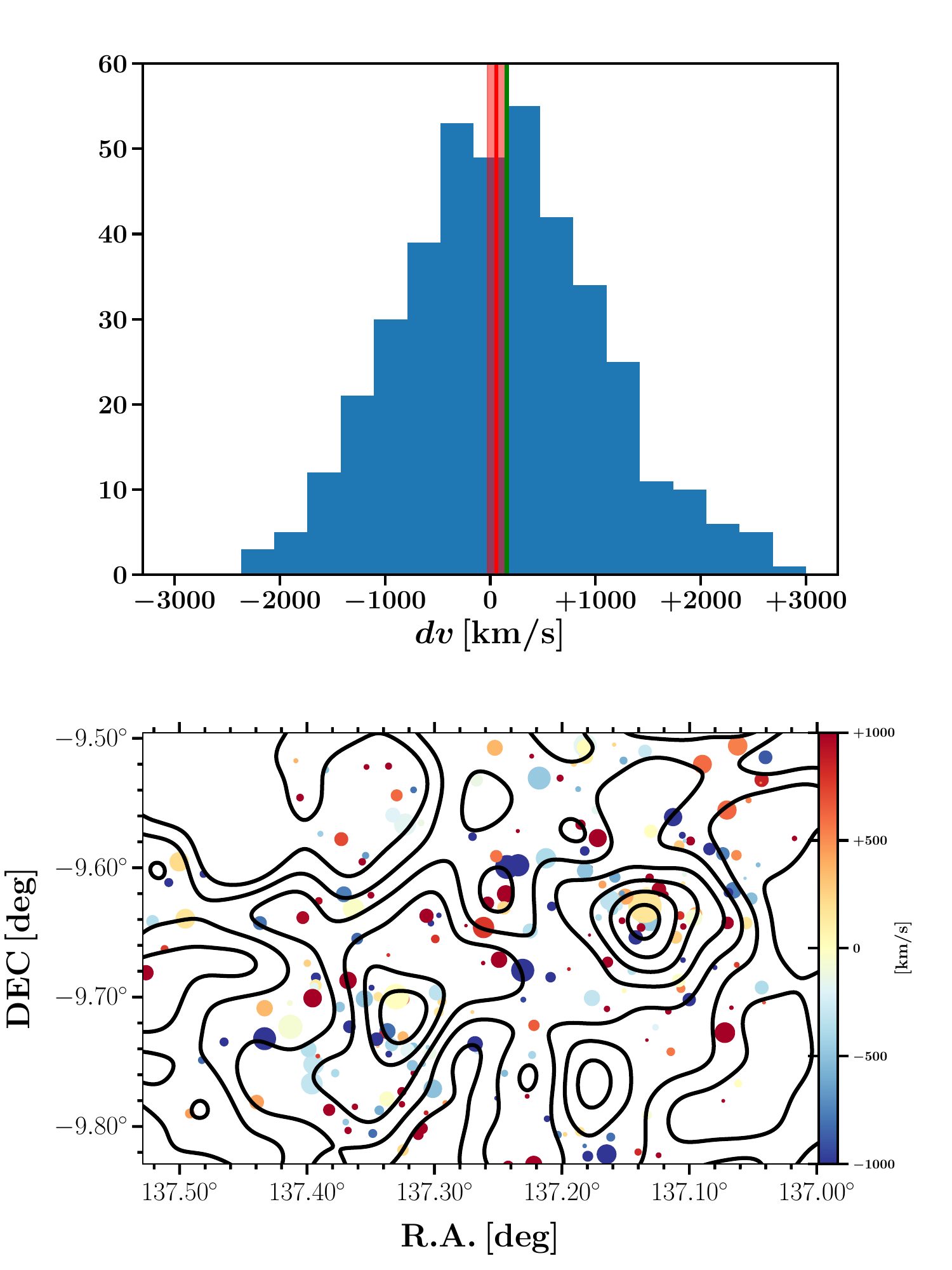}
    \caption{{\it Top}: The velocity histogram relative to the E-BCG. The vertical red line and the shaded area indicate the relative velocity from $z_{\rm cbi}$ and its uncertainty, respectively. The vertical green line shows the relative velocity of the W-BCG. {\it Bottom}: The scatter plot of the line-of-sight velocities of the member galaxies. Red and blue colors represent positive and negative line-of-sight velocities relative to the main cluster, respectively. The circle size is proportional to the square root of the luminosity. The contours indicate the significance levels of the WL mass distribution, as in Figure \ref{fig:massmap}. {Altext: Two figures aligned vertically.}}
    \label{fig:vmap}
\end{figure}

We retrieved spectroscopic redshifts of galaxies associated with A754 from the NASA/IPAC Extragalactic Database (NED). Duplicate entries were removed by cross-matching objects within $3\,$arcsec, and two additional galaxies whose positions do not align with those in the HSC image were excluded. We then iteratively calculated the central biweight redshift, $z_{\rm cbi}$, of the member galaxies. Member candidates were selected as those lying within a radius of $2\,h_{70}^{-1}\,\mathrm{Mpc}\simeq r_{200}^{\rm tot}$ from the E-BCG and having a relative velocity $v=(z-z_{\rm cbi})c/(1+z_{\rm cbi})$ that falls within twice the expected caustic amplitude estimated from the total WL mass. A total of 401 spectroscopic galaxies is selected. The resulting redshift is $z_{\rm cbi}=0.054487_{-0.000282}^{+0.000302}$, which agrees with $z_{\rm EB}=0.054291$ within the $1\sigma$ uncertainty. When we repeated the calculation at intervals of $1\,h_{70}^{-1}\,\mathrm{Mpc}$ out to $3\,h_{70}^{-1}\,\mathrm{Mpc}$, the results showed no significant variation. Therefore, we adopt $z_{\rm EB}$ as the redshift of the galaxy cluster (top panel of Figure~\ref{fig:vmap}). The difference between $z_{\rm cbi}$ and $z_{\rm EB}$ is negligible for the WL mass measurement.

The luminosity map for the spectroscopically identified galaxies is shown in the bottom panel of Figure \ref{fig:massmap}, revealing a clear double-peak structure. The W-BCG is more luminous than the E-BCG, although the western component is less massive than the eastern one. This is consistent with the large intrinsic scatter in BCG stellar masses ($0.70_{-0.05}^{+0.06}$) reported by \citet{2022PASJ...74..175A}.

The relative velocity map is shown in the bottom panel of Figure \ref{fig:vmap}. No spatial variation in the velocity structure is visually discernible. We also attempted to identify local velocity structures related to the WL and luminosity substructure utilizing a Gaussian-mixture model (GMM), but did not succeed. Instead, we estimated $z_{\rm cbi}$ within a scale radius ($0.21\,h_{70}^{-1}\,{\rm Mpc}$) of the subcluster from the W-BCG and obtained $0.054154_{-0.000670}^{+0.001813}$, which is in agreement with $z_{\rm WB}$ within $1\sigma$. When the radius is varied to $0.3\,h_{70}^{-1}\,{\rm Mpc}$ or $0.5\,h_{70}^{-1}\,{\rm Mpc}$, the result remains unchanged. Therefore, we consider $z_{\rm WB}$ as the subcluster redshift in the following subsection.

\subsection{Merger geometry} \label{subsec:dis}

\subsubsection{Model}
One advantage of WL mass measurements is their ability to help us investigate the parameters involved in cluster mergers, reducing the time to run numerical simulations.  We here searched for the merger parameters by calculating the motion of each cluster within the framework of the two-body problem of a point mass, incorporating Chandrasekhar's dynamical friction \citep{1943ApJ....97..255C} with a modification.
Since the dynamical friction of the massive subcluster is not trivial (Appendix \ref{app:badmodel}), we calibrate it by comparing with numerical simulations from the Galaxy Cluster Merger Catalog website \citep{2011ApJ...728...54Z,2018ApJS..234....4Z}. The simulations include the dark matter and the ICM components, and the secondary effects of tidal deformation, non-uniform dynamical friction, and so on.
The following approximate formula allows us to characterize situations even with major merger cases; 
\begin{eqnarray}
    \ddot{\bm r}_a &=&-\frac{GM_b(<r) \bm{r}}{r^3}+\bm{F}_{{\rm dyn},ab},\label{eq:trajectory} \\
    \bm{F}_{{\rm dyn},ab}&=& -4\pi G^2 M_a(<r) \rho_{b}(r)\ln \Lambda (r) A_{{\rm vel},ab}\frac{\bm{v_{ab}}}{v_{ab}^3} ,\nonumber \\
    A_{{\rm vel},ab}&=&{\rm erf}(X_{ab})-\frac{2X_{ab}}{\sqrt{\pi}}e^{-X_{ab}^2}, \quad X_{ab}=v_{ab}/(\sqrt{2} \sigma_{v,b}),\nonumber 
    \end{eqnarray}
    for
    \begin{eqnarray}
    \ln\Lambda&=& \begin{cases}
    \ln \left(1+\frac{r^2}{p_{90}^2}\right)^{1/2}    & \hspace{-6em} (r\ge p_{90}=\frac{GM_a}{v_{ab,{\rm ini}}^2},M_a < M_b) \\
    \ln \left(1+\frac{r^2}{p_{90}^2}\right)^{1/4}\left(1+\frac{r^2}{p^2}\right)^{1/4} & \\
    +  \ln\left(1+\frac{M_b(<r)}{M_a(<r)}\right)&  ({\rm else})\\
      \end{cases}  
\end{eqnarray}
We calculated the trajectories of the main and subclusters by swapping the subscripts, $a$ and $b$, where $\bm{r}_a$ represents the position of component $a$, and $\bm{v}_{ab}$ is the velocity of $a$ relative to a background component $b$, which has a mass density given by $\rho_b$. Meanwhile, $\sigma_{v,b}$ indicates the velocity dispersion derived from the mass $M_b(<r)$ through the Jeans equation under the assumption of isotropic dispersion.

\begin{figure}[!ht]
    \centering
    \includegraphics[width=\linewidth]{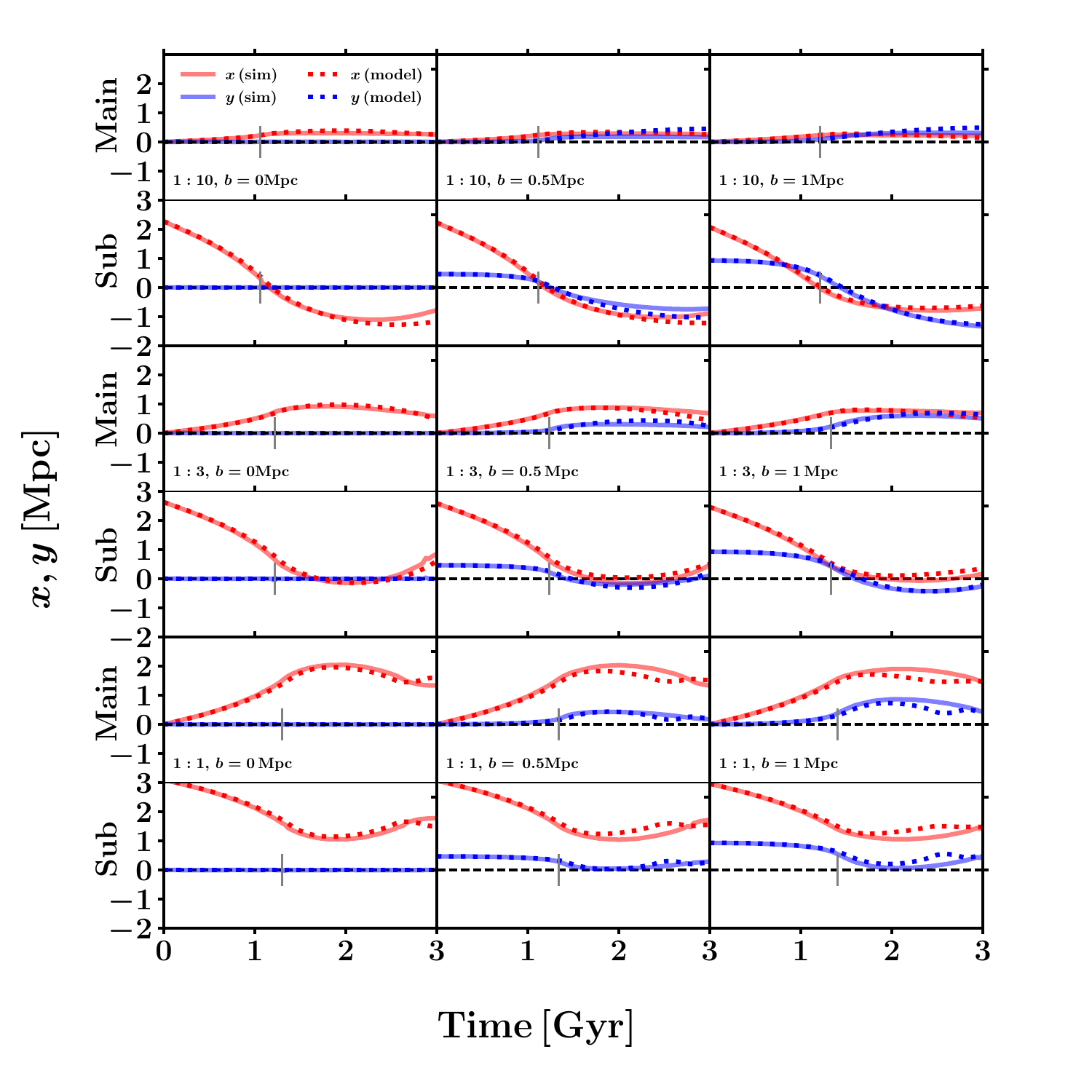}
    \flushleft
    \caption{Trajectories of the main and subclusters versus the merger time. The $x$ and $y$ positions obtained from numerical simulations \citep{2011ApJ...728...54Z,2018ApJS..234....4Z} are represented by the solid red and blue lines, respectively. The dotted red and blue lines correspond to $x$ and $y$ of the analytical model as described by equation (\ref{eq:trajectory}), respectively. The vertical gray lines represent the pericenter time.
     Results from the main cluster are presented in the first, third, and fifth rows, while those from the sub-cluster appear in the second, fourth, and sixth rows. The inset annotations denote the parameters utilized for merging. {Altext: a panel layout consisting of 3 columns and 6 rows.}
    }
    \label{fig:trajectorymodel}
\end{figure}

A Coulomb logarithm, denoted as $\ln \Lambda$, is expressed by $\ln \left(1 + \frac{r^2}{p_{90}^2}\right)^{1/2}$, where $p_{90}$ represents the impact parameter associated with a 90-degree deflection \citep{2008gady.book.....B}. When $r/p_{90}<1$ of the subcluster, the model trajectories for both the main and subclusters are depicted better by a modified Coulomb logarithm combined with the initial parameter ($p_{90}$) and the local deflection term ($p = \frac{GM_a(<r_a)}{v^2_{ab}}$), and a conventional term \citep[e.g.][]{2016MNRAS.463..858P} that indicates a mass-dependent interaction efficiency. We ignored the dynamic friction due to the non-uniform distribution, tidal deformation, and a change in the concentration parameter. Employing this formula, 
the positions of the analytical solution remain accurate to within approximately $0.1-0.3$ Mpc for up to 3 Gyr in simulations featuring mass ratios of $1/1$, $1/3$, and $1/10$, along with impact parameters of $b=0, 0.5$, and $1$ Mpc, as shown in Figure \ref{fig:trajectorymodel}. Because velocities in the public data are available only for the ICM and not for the mass, a direct comparison of velocities is not presented here. Initial velocities can be estimated from the setup parameters described in \citet{2011ApJ...728...54Z} and $M_{200}$ of the main and sub clusters. For cases with mass ratios of 1:10 with $b=0.5$ Mpc, 1:1 with $b=0.5$ Mpc, and 1:1 with $b=1$ Mpc, the discrepancy between the analytical model and the simulations is somewhat larger. Developing a more accurate approximation formula for major mergers over longer time scales remains an open problem for the community.

\begin{figure}
    %\centering
    \includegraphics[width=\linewidth]{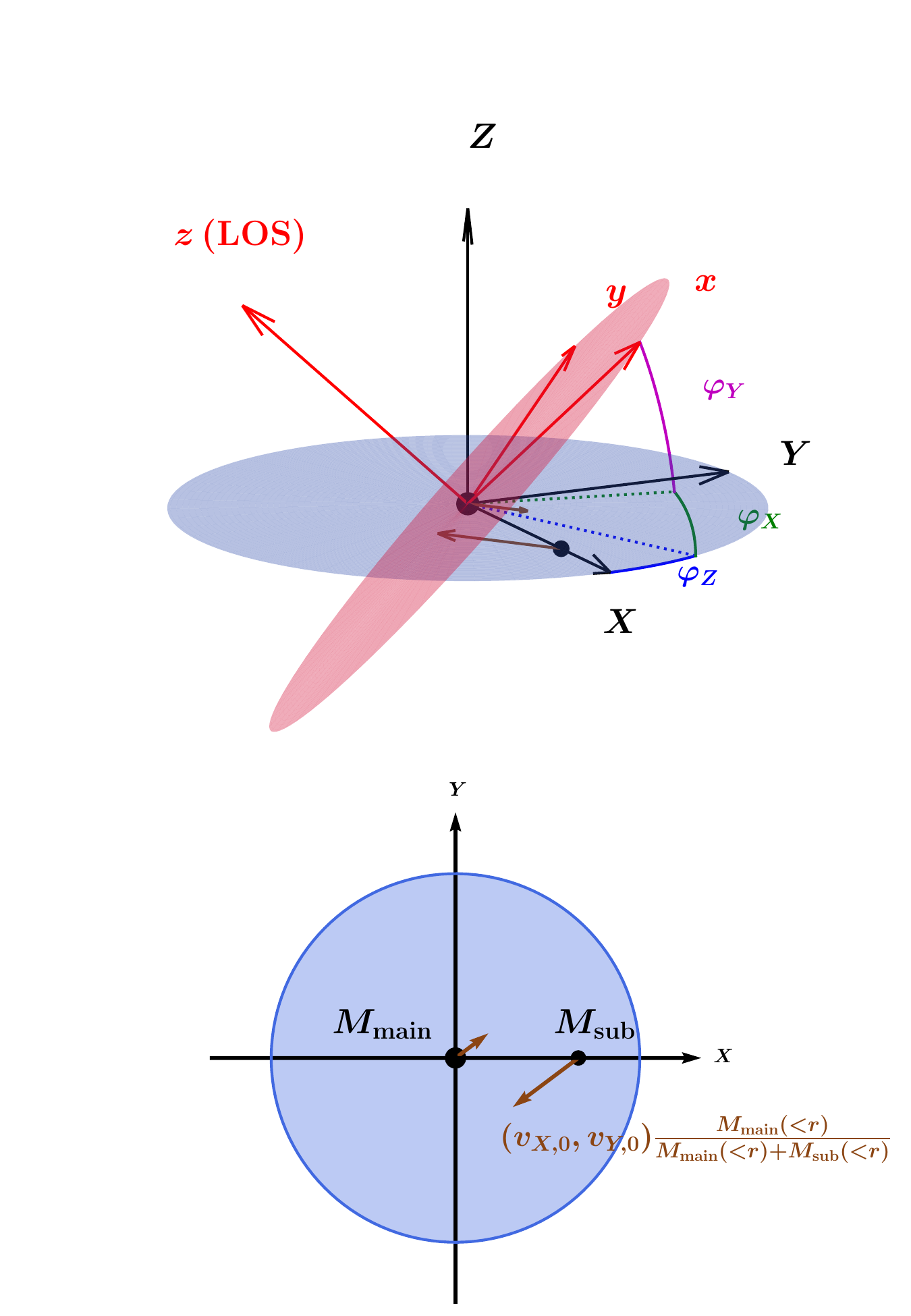}
    \caption{Illustration of the merger configuration. {\it Top}: The blue and red disks indicate the merger planes before and after the Euler rotation, respectively, and the solid black and red arrows represent the coordinate axes in each state. The blue, green, and magenta curves show the Euler angles of $\varphi_Z,\varphi_X$, and $\varphi_Y$. The blue and green dotted lines trace the location of the $X$ axis after the rotations $R_Z(\varphi_Z)$ and $R_X(\varphi_X)R_Z(\varphi_Z)$, respectively. The main and sub clusters are marked by large and small points, respectively. The brown arrows depict the velocity vectors of the main and sub clusters. {\it Bottom}: View of the merger plane along the $Z$ axis. The definitions of the symbols are identical to those in the top panel. {Altext: Two figures aligned vertically.}
    }
    \label{fig:param}
\end{figure}

We first set up the merger plane ($X, Y$), placing the main cluster at the origin in the initial state. Since we assume a spherically symmetric mass distribution, motions perpendicular to the merger plane do not need to be considered. We adopt an initial position of the subcluster at $(2,0)$ Mpc and treat the relative initial velocity, $v_{X,0}$ and $v_{Y,0}$, as free parameters. The initial velocity of the subcluster is calculated by $({v}_{X,0},{v}_{Y,0})M_{\rm main}(<r)/(M_{\rm main}(<r)+M_{\rm sub}(<r))$. 
The radial infall of the subcluster toward the main cluster is represented by allowing $v_{X,0}$ to take negative values. The initial impact parameter can then be estimated from these initial velocities. Finally, we adjust the orientation of the merger plane using Euler angles to match the observational plane of the sky, 
\begin{eqnarray}
    {\bm x}=R_Y(\varphi_Y)R_X(\varphi_X)R_Z(\varphi_Z){\bm X}.
\end{eqnarray}
Here, ${\bm x}=(x,y,z)$ denotes the coordinate system in which the positive $x$, $y$, and $z$ axes point toward the west, north, and away from the observer along the line-of-sight, respectively. The Euler angle, $\varphi_Z$, changes the rotation of the merger plane. The inclination angle between the line-of-sight and the merger plane is given by $\arcsin(|\cos(\varphi_X)\cos(\varphi_Y)|)$. The set of free dynamical parameters is ${\bm p}_{\rm dyn}=(v_{X,0},v_{Y,0},\varphi_X,\varphi_Y,\varphi_Z)$. The merger parameters are illustrated in Figure \ref{fig:param}.

\subsubsection{Trajectory determination}

As described in Sec \ref{subsec:specz}, we considered the redshifts of W-BCG and E-BCG to be indicative of the main and subclusters, respectively, namely, the line-of-sight velocity of the subcluster of $\tilde{v}_{{\rm WB},z}=+156\pm153\,{\rm km\,s^{-1}}$. Hereafter, we denote the relative position and velocity of the subcluster with respect to the main cluster by the symbols $\tilde{{\bm x}}_{\rm sub}$ and $\tilde{{\bm v}}_{\rm sub}$, respectively. The measurement error was estimated from 10,000 bootstrapping iterations within a radius of $0.7\,h_{70}^{-1}$ Mpc (slightly smaller than the separation of two BCGs) around each BCG, allowing overlaps, with the standard deviation of the velocity differences derived from $z_{\rm cbi}$ in each iteration adopted as the corresponding error. 
The radius was chosen to maximize the number of galaxies while excluding the other BCG.
Given that gravitational forces bind the main and subclusters, the effects of the Hubble flow and the cosmic expansion are disregarded throughout the merging process.

The hydrodynamic interaction induced by cluster mergers is dominated around the pericenter because the ram pressure of the highest velocity in a dense ICM environment and the rotation of the tidal spin are significant. Hence, the ICM core associated with the subcluster that rotates by tidal force could be stripped away by ram pressure. Once the stripped gas is free from this rotation, the gas moves inertially in alignment with the tangential velocity at the pericenter \citep{2019ApJ...874..112S,You2025}. In simulations, the filament induced by the interaction between the main and subclusters during the pericenter phase remains present to some extent even after apocenter passage; however, because the formed filament is pulled back by the gravitational potential of the main cluster, significant changes in the position of the main cluster can alter its original structure. In this analysis, we assume the scenario before apocenter.

Therefore, the major axis of the filamentary structure tends to be parallel to the tangential (moving) direction at the pericenter (Appendix \ref{app:sim}). We express the tangential angle in the sky plane at the pericenter as $\phi_{\rm peri}=\arctan({\tilde v}_y/{\tilde v}_x)(t_{\rm peri})$, where the angle is measured from west to north.

Given the analytical model (eq. \ref{eq:trajectory}), the likelihood for the merger trajectory is formulated as follows:
\begin{eqnarray}
    p( {\bm d} | {\bm p}_{\rm dyn},{\bm p} ) &\propto& \exp\left[-\frac{1}{2}\left(\frac{({\tilde v}_z(t_0)-{\tilde v}_{{\rm WB},z}^{\rm opt})^2}{\sigma_{\rm los}^2}\right.\right. \label{eq:LikelihoodDyn} \\
   &&\hspace{-7em}+ \left.\left.\frac{({\tilde x}_{\rm sub}(t_0)-{\tilde x}_{\rm sub}^{\rm WL})^2+({\tilde y}_{\rm sub}(t_0)-{\tilde y}_{\rm sub}^{\rm WL})^2}{\sigma_{\rm pos}^2}+\frac{(\phi_{\rm peri}-\phi_{\rm fil}^{\rm X})^2}{\sigma_{\phi}^2}\right)\right]. \nonumber
\end{eqnarray}
Here, ${\bm d}$ denotes the data array, $\sigma_{\rm pos}=0.046\,h_{70}^{-1}{\rm Mpc}$ is the typical uncertainties of the WL centroid determination,  ${\tilde {\bm x}}_{\rm sub}^{\rm WL}$ is the relative position of the subcluster determined by the 2D WL analysis, $t_0$ represents the present time, determined by a minimum value of the sum of the first and second terms. The orientation angle of the filament is measured by the X-ray image; $\phi_{\rm fil}^{\rm X}=16.4\pm0.5$ deg. Here, we first made a smoothed Gaussian image with a FWHM of $=2\sqrt{2\ln2}\times 1'\simeq 2.'3$, which was used as a mask to select the filamentary region based on the minimum and maximum count-rate thresholds to exclude the low X-ray surface brightness region and the eastern tadpole-shaped gas (the top panel of Figure \ref{fig:massmap}). We then performed an orthogonal regression using Singular Value Decomposition (SVD), weighted by the count rate of the non-smoothed image, for different combinations of smoothed count-rate intervals. We adopted its average and standard deviation of the angles as the orientation angle of the filament and its uncertainty. The line-of-sight velocity of the optical W-BCG galaxy is expressed with a superscript of opt, and its uncertainty is denoted by $\sigma_{\rm los}$.

The recent {\it XRISM} observations \citep{2025arXiv251016553O} revealed that the line-of-sight motions of the middle of the filamentary gas and the tadpole-shaped gas core (the top panel of Figure \ref{fig:massmap}) relative to the E-BCG are $-436^{+26}_{-28}\,{\rm km\,s^{-1}}$ and $+220^{+26}_{-29}\,{\rm km\,s^{-1}}$, respectively. The tidal torque acts to align the tidal spin with the orbital motion, because the tidal bulge is slightly offset from the line connecting the two bodies, producing a torque that drives the rotation toward the moving direction (Appendix \ref{app:sim}).
Consequently, the azimuthal movement of the subcluster within the $\tilde{x}$-$\tilde{z}$ plane is anticipated to occur in an anti-clockwise trajectory. Given the above likelihood, we impose the additional constraint of $p({\tilde v}_z)p({\tilde x})=\mathcal{H}({\tilde v}_z(t=0))\mathcal{H}(-{\tilde x}(t=0))$.
Moreover, at the northern part of the tadpole-like gaseous structure, it is displaced from the central mass peak. This offset could be induced by the subcluster's transit in the northern region, which is described by $p({\tilde y})={\mathcal H}({\tilde y}(t_{\rm peri}))$.

We first fixed the best-fit WL parameters to determine the merger trajectory with a posterior distribution of $ p( {\bm d} | {\bm p}_{\rm dyn},{\bm p} )p({\bm {\tilde v}})p({\bm {\tilde x}})p({\bm p}_{\rm dyn})$. A flat prior distribution was adopted for the dynamics parameters ($p({\bm p}_{\rm dyn})$), specifically within the constraints $-2400\,{\rm km\,s^{-1}}< v_{X,0} < -1000\,{\rm km\,s^{-1}}$, $0\,{\rm km\,s^{-1}} < v_{Y,0} < 1200\,{\rm km\,s^{-1}}$, $-90^\circ \le \varphi_X \le 90^\circ$, $0^\circ \le \varphi_Y < 360^\circ$, and $0^\circ \le \varphi_Z < 360^\circ$.
The resulting parameters are shown in Table \ref{tab:dyn}. The parameters of $v_{Y,0}$ cannot be constrained well. 
We next selected the WL parameters within their $1\sigma$ uncertainties from the posterior distributions at each step of the MCMC run, in order to account for the uncertainties in the WL masses when estimating the trajectory. We found that the result remains consistent with the first result (Table~\ref{tab:dyn}). Therefore, the discussion that follows is based on the first result. We also attempted a joint fit using both the 2D shear pattern and the dynamical constraints, but the acceptance rate was low; consequently, the joint-fit results are not included in this paper.

\begin{table}[]
    \centering
    \begin{tabular}{c|cc}
  case  & 1 & 2 \\ 
    \hline
 $M_{200}^W$        & $(3.13)$  & $(3.13_{-1.00}^{+1.53})$\\
  $c_{200}^W$  & $(6.41)$ & $(6.41_{-1.88}^{+3.02})$ \\
$\alpha^W$ & $(137.136)$ & $(137.136_{-0.004}^{+0.005})$ \\
 $\delta^W$  & $(-9.638)$ & $(-9.638_{-0.005}^{+0.006})$  \\
 $M_{200}^E$     & $(6.41)$  & $(6.41_{-1.97}^{+2.92})$  \\
$c_{200}^E$  & $(2.71)$& $(2.71_{-0.88}^{+1.20})$ \\
$\alpha^E$ & $(137.325)$ & $(137.325_{-0.012}^{+0.011})$\\
 $\delta^E$ & $(-9.712)$ & $(-9.712_{-0.026}^{+0.010})$ \\
 $v_{X,0}$ &  $-1665.29_{-410.59}^{+330.88}$ & $-1811.64_{-346.24}^{+403.75}$\\
  $v_{Y,0}$ & $699.61_{-395.28}^{+320.27}$&  $904.13_{-438.29}^{+182.96}$\\
  $\varphi_{X}$  & $-68.84_{-5.90}^{+18.53}$  &  $-67.85_{-6.86}^{+22.26}$\\
  $\varphi_{Y}$ & $195.41_{-46.80}^{+58.39}$ &  $183.15_{-20.52}^{+34.88}$\\
  $\varphi_{Z}$  & $303.36_{-28.89}^{+24.11}$  &   $311.66_{-34.98}^{+17.16}$\\
    \end{tabular}
    \flushleft
    \caption{The results of dynamical analysis. The units of the masses, the velocities, the positions, and the Euler angles are $10^{14}h_{70}^{-1}M_\odot$, km s$^{-1}$, degree, and degree, respectively. The round brackets are the WL mass parameters used in the analysis.}
    \label{tab:dyn}
\end{table}

\begin{figure}
    \includegraphics[width=\linewidth]{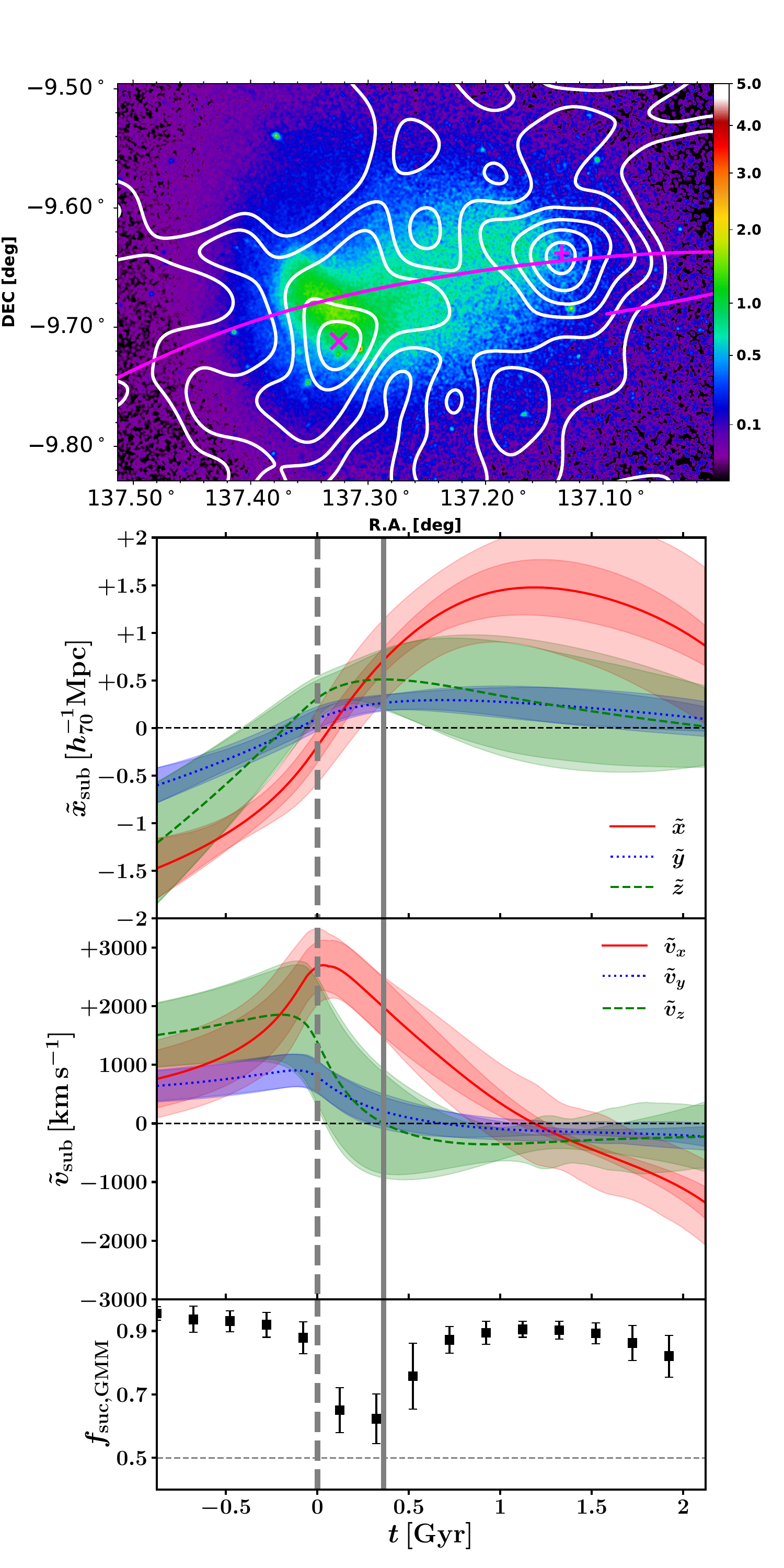}
    \caption{{\it Top}: the magenta line represents the merger trajectory in the rest frame of the main cluster, overlaid on the top panel of Figure \ref{fig:massmap}. The magenta $+$ and $\times$ are the NFW centers determined by 2D WL analysis.
    {\it Middle} : the relative position of the subcluster versus the merger time. The time is set to be zero at the pericenter. The red solid, blue dotted, and green dashed lines are the ${\tilde x}$, ${\tilde y}$, and ${\tilde z}$ positions, respectively. The dark and light shaded regions indicate the $1\sigma$ uncertainty ranges, evaluated first using only the Euler angles and then using the full set of parameters. The vertical gray line expresses the current time. The vertical gray dashed line is the pericenter.  {\it Second-to-bottom}: the relative velocity of the subcluster versus the merger time. {\it Bottom}: the successful ratio, $f_{\rm suc,GMM}$, to identify member galaxies in mock spec-z data using the GMM method. The horizontal line is the probability of the successful ratio by chance. {Alttext: fourth vertically aligned figures.}}
    \label{fig:trajectory}
\end{figure}

\begin{figure}
    \centering
    \includegraphics[width=\linewidth]{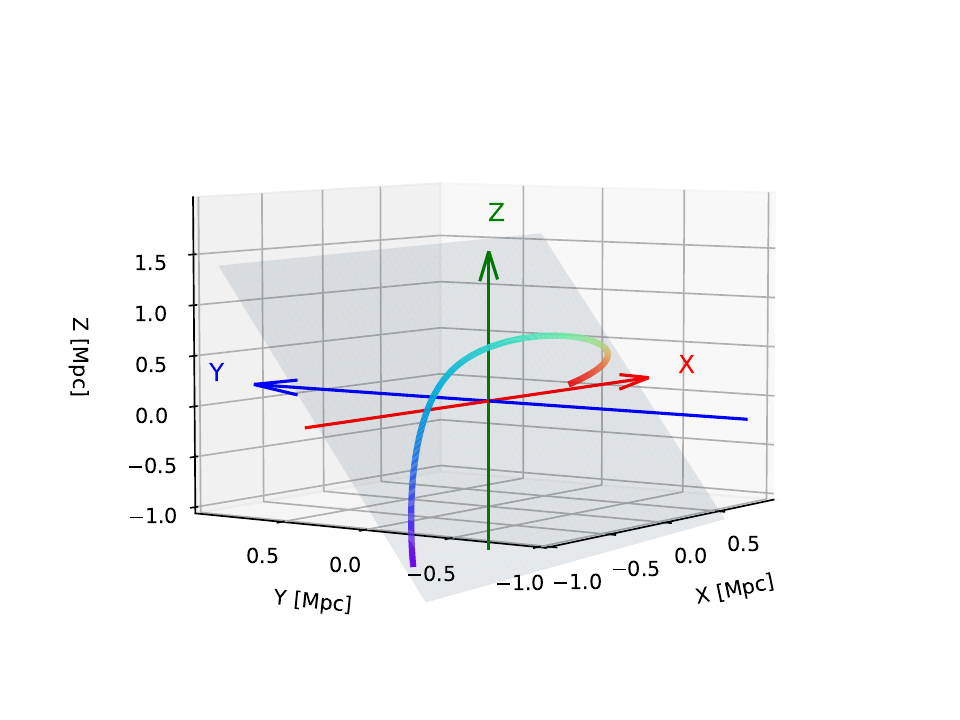}
    \flushleft
    \caption{The trajectory in the three-dimensional space. The transparent blue plane denotes the merger plane. The colors along the trajectory curve indicate the progression of time, from blue to red. {Altext:three-dimensional figure}}
    \label{fig:3Dtrajectory}
\end{figure}

The best-fit merger paths are illustrated in Figure \ref{fig:trajectory}. At an initial separation of 2 $h_{70}^{-1}$ Mpc, the impact parameter and the initial velocity are $b=0.77\pm0.36\,h_{70}^{-1}\,{\rm Mpc}$ and $v_0= 1806\pm319\,{\rm km\,s^{-1}}$, indicating an off-axis merger. The angle of inclination between the line-of-sight direction and the merger plane is $20\pm19$ deg, indicating a near-line-of-sight merger. This is unexpected, as we initially hypothesized that the cluster represented a merger in the plane of the sky because of the lack of a distinct double peak configuration in redshift space. This is simply due to the following reason.
The direction of motion within the merger plane changes after the pericenter passage when the impact parameter is finite (middle and second-bottom panels of Figure \ref{fig:trajectory}). This change arises because the gravitational interaction between the two clusters deflects the subcluster’s trajectory. After passing through the pericenter, the tangential component of the motion becomes dominant and the subcluster appears to move primarily along the plane of the sky. Figure \ref{fig:3Dtrajectory} provides a depiction of the trajectory within a three-dimensional space, facilitating comprehension of its path.

Approximately $0.88\pm0.09$ Gyr after the starting time, the pericenter is reached, with the separation distance of $0.38\pm0.22\,h_{70}^{-1}\,{\rm Mpc}$ and a relative velocity of $3114\pm420\,\mathrm{km\,s^{-1}}$, predominantly consisting of a tangential velocity. Notably, this magnitude surpasses the bulk velocity of the main gas as measured by {\it XRISM} observations. The uncertainties were quantified by generating 5000 resampled realizations of the posterior distribution for each orbit and determining the corresponding pericenter for each realization. The velocity component along the $z$-axis at the pericenter is approximately $1383\pm 1160\,{\rm km\,s^{-1}}$. When the orientation angles are taken into account, uncertainties in the Euler angles propagate into and substantially increase the uncertainty in the velocity vector in the observational frame. As shown in Figure~\ref{fig:trajectory}, the Euler angles are the dominant source of orbital errors for $\tilde{y},\tilde{z},\tilde{v}_y$, and $\tilde{v}_z$. The uncertainties in the initial velocity is the predominant source of error in $\tilde{x}_{\rm sub}$ after around the pericenter and $\tilde{v}_{x,\rm sub}$ after $t\simgt 0.5$ Gyr.

The present time is  $\sim 0.36\pm0.24$ Gyr after pericenter passage. This time is determined by minimizing the sum of the first and second terms in eq. (\ref{eq:LikelihoodDyn}).
The three-dimensional distance between the main and subclusters is about $0.9\pm0.27\,h_{70}^{-1}{\rm Mpc}$ and the position of the line-of-sight is ${\tilde z}_{\rm sub}\simeq 0.5\pm0.6\,h_{70}^{-1}{\rm Mpc}$. The radial and tangential velocities are $\simeq1600\pm850\,{\rm km\,s^{-1}}$ and $\simeq1183\pm501\,{\rm km\,s^{-1}}$, respectively.

After the pericenter, the line-of-sight velocity becomes a low and constant value, making it difficult to distinguish between the main and subclusters using only spectroscopic redshift data. To verify this, we performed the following simple simulation. We generated 200 main-cluster galaxies and 100 sub-cluster galaxies, which are randomly distributed according to each NFW profile within $r_{200}$. We then estimated the redshift using the merger motions and cluster velocity dispersions as a function of the merger time. We tried to identify which galaxies belong to the main and subclusters by employing the GMM method $50$ times (the bottom panel of Figure \ref{fig:trajectory}.
The probability of accurately identifying galaxies ($f_{\rm suc, GMM}$) is notably high, exceeding 0.9, before reaching the pericenter, whereas it decreases drastically to $\sim 0.5-0.7$ just after pericenter. Given that a success rate of 0.5 occurs by chance, this suggests significant difficulty in distinguishing member galaxies at the current stage, which agrees well with our results. With the passage of more time, the increased spatial separation permits a clear identification of individual galaxies within the cluster, leading to an increase in $f_{\rm suc, GMM}$.

The dynamics of cluster mergers cannot be fully characterized solely on the basis of spectroscopic redshift data of the member galaxies. In the absence of constraints on a change of the motion by a finite impact parameter, the merger trajectory may be misinterpreted.
Using an analytical model, we can reconstruct the three-dimensional motion by combining 2D WL, spectroscopic redshift, and X-ray information. In reality, several additional factors, such as halo triaxiality, tidal deformation, tidal rotation, tidal destruction, and non-uniform dynamical friction, should be considered, which will be considered in numerical simulations.
However, our simplified approach could reduce the computational time of numerical simulations for parameter search to realize the merger situation, including hydrodynamical effects and {\it XRISM} observations.

\section{Summary} \label{sec:summary}

We carried out 2D WL analysis using the Subaru/HSC and Suprime-Cam data, taking into account the uncorrelated LSS lensing effect. We found that the eastern and western mass components are the main and subclusters, respectively. 
The mass ratio of the cluster is $f_{\rm ratio}=0.49\pm0.33$, signifying a major merger. Here, the contribution of LSS lensing constitutes approximately 50\% of the total uncertainty.

We developed an approximate framework to compute the trajectories of major mergers, treating the system as a two-body problem of a point mass with a modified Chandrasekhar dynamical-friction term and incorporating Euler angles. Since the dynamical friction of the major mergers is not trivial, we calibrate it by comparing with numerical simulations from the Galaxy Cluster Merger Catalog \citep{2011ApJ...728...54Z,2018ApJS..234....4Z}. 

Given this framework, we determined the merger trajectories by combining the 2D WL measurements, archival spectroscopic redshifts of member galaxies, the orientation angle of the X-ray filament, and priors based on X-ray kinematics. We found that the impact parameter and initial velocity are $b \simeq 0.77\pm0.36\,h_{70}^{-1}\,{\rm Mpc}$ and $v_0 \simeq 1806\pm319\,{\rm km\,s^{-1}}$ at an initial separation of $2\,h_{70}^{-1}$ Mpc, indicating an off-axis merger. The inclination angle between the line-of-sight direction and the merger plane is $20 \pm 19$ degrees. Thus, the merger occurs on a plane that is only slightly inclined with respect to the line of sight.

This highlights that the absence of a clear structure in galaxy redshift space does not necessarily imply that the merger occurs strictly on the plane of the sky. The apparent lack of line-of-sight motion can simply arise from a change in the direction of motion within the merger plane after core passage. Our method provides an efficient way to explore merger parameters that reproduce the hydrodynamic features induced by cluster mergers and the conditions probed by {\it XRISM} observations through numerical simulations.

\section*{Acknowledgements}

We are grateful for the reviewer’s constructive and insightful comments.

This paper is based [in part] on data collected at the Subaru Telescope and retrieved from the HSC data archive system, which is operated by Subaru Telescope and Astronomy Data Center (ADC) at NAOJ. Data analysis was in part carried out with the cooperation of Center for Computational Astrophysics (CfCA), NAOJ. We are honored and grateful for the opportunity of observing the Universe from Maunakea, which has the cultural, historical and natural
significance in Hawaii.

This work was supported by JSPS KAKENHI grant numbers 
JP25K07368 (N.O.), JP20H00157 (K.N.), and JP25K23398 (S.U.).
Y.O. would like to take this opportunity to thank the ``Nagoya University Interdisciplinary Frontier Fellowship'' supported by Nagoya University and JST, the establishment of university fellowships towards the creation of science technology innovation, Grant Number JPMJFS2120.
Y.O. was supported by the Sasakawa Scientific Research Grant from The Japan Science Society.
N.O. acknowledges partial support by the Organization for the Promotion of Gender Equality at Nara Women's University.
S.U. acknowledges support by Program for Forming Japan's Peak Research Universities (J-PEAKS) Grant Number JPJS00420230006.

This work made use of data from the Galaxy Cluster Merger Catalog (http://gcmc.hub.yt).

\appendix

\section{Unfavorable Trajectory Models} \label{app:badmodel}

\begin{figure}
    \centering
    \includegraphics[width=\linewidth]{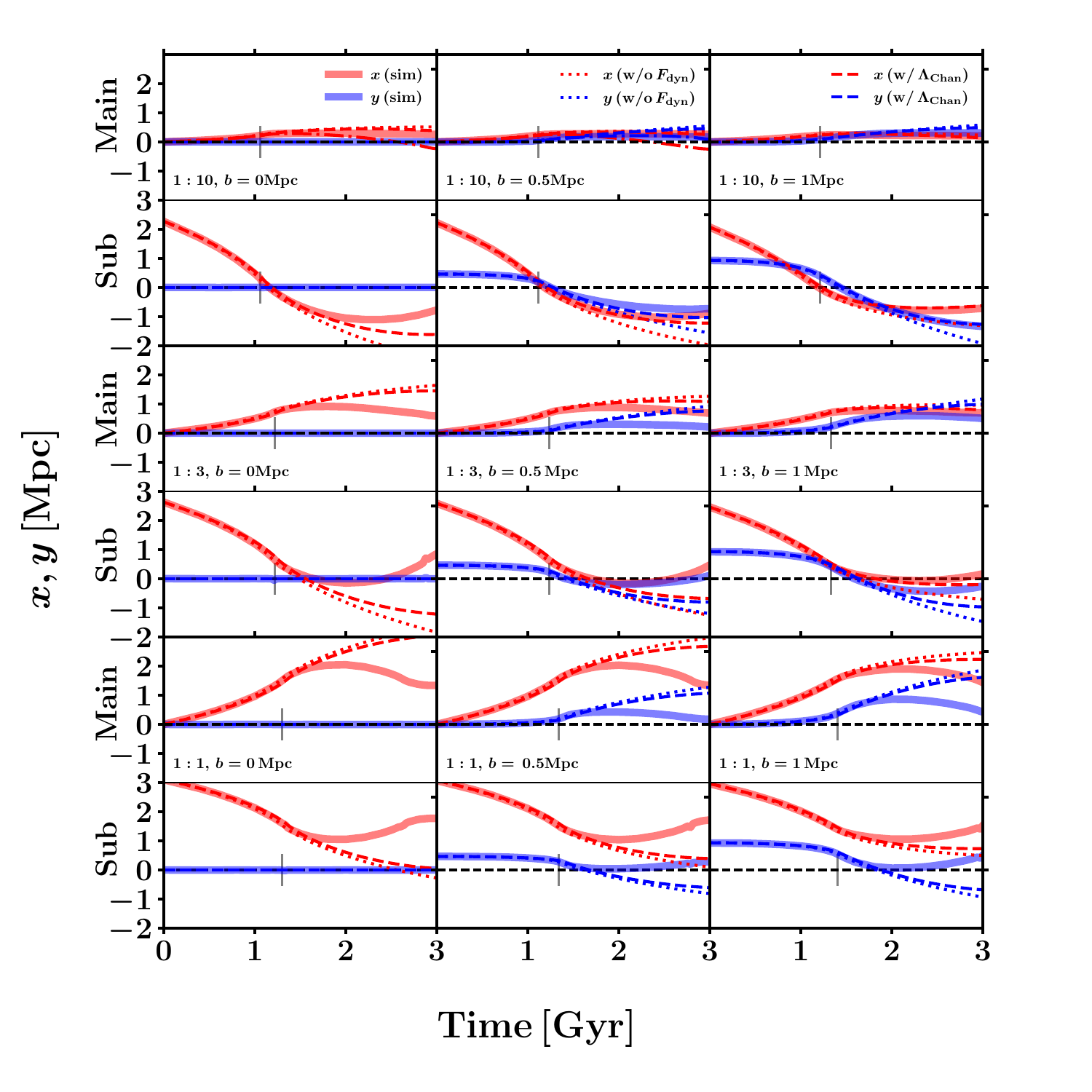}
    \flushleft
    \caption{Trajectories of the main and subclusters versus the merger time. The same figure as Figure \ref{fig:trajectorymodel}, except for the analytical model. The dotted lines are a trajectory without dynamical friction. The dashed lines are the trajectories with the classical Chandrasekhar's dynamical friction \citep{1943ApJ....97..255C}. {Altext: a panel layout consisting of 3 columns and 6 rows.}}
    \label{fig:badmodel}
\end{figure}

We also examined several alternative trajectory models that fail to reproduce the numerical simulations.
Figure \ref{fig:badmodel} illustrates the two trajectories among the failed models. The trajectories (dotted lines) without the dynamical friction are unsuccessful in all the cases. The classic Chandrasekhar dynamical friction model \citep[dashed line;][]{1943ApJ....97..255C} is effective solely for mass ratios of $1:10$ when $b=1$ Mpc, as this configuration meets the criterion $r>p_{90}$; however, it is inadequate for other cases.

\section{X-ray filamentary structure in numerical simulations} \label{app:sim}

\begin{figure}
    \centering
    \includegraphics[width=\linewidth]{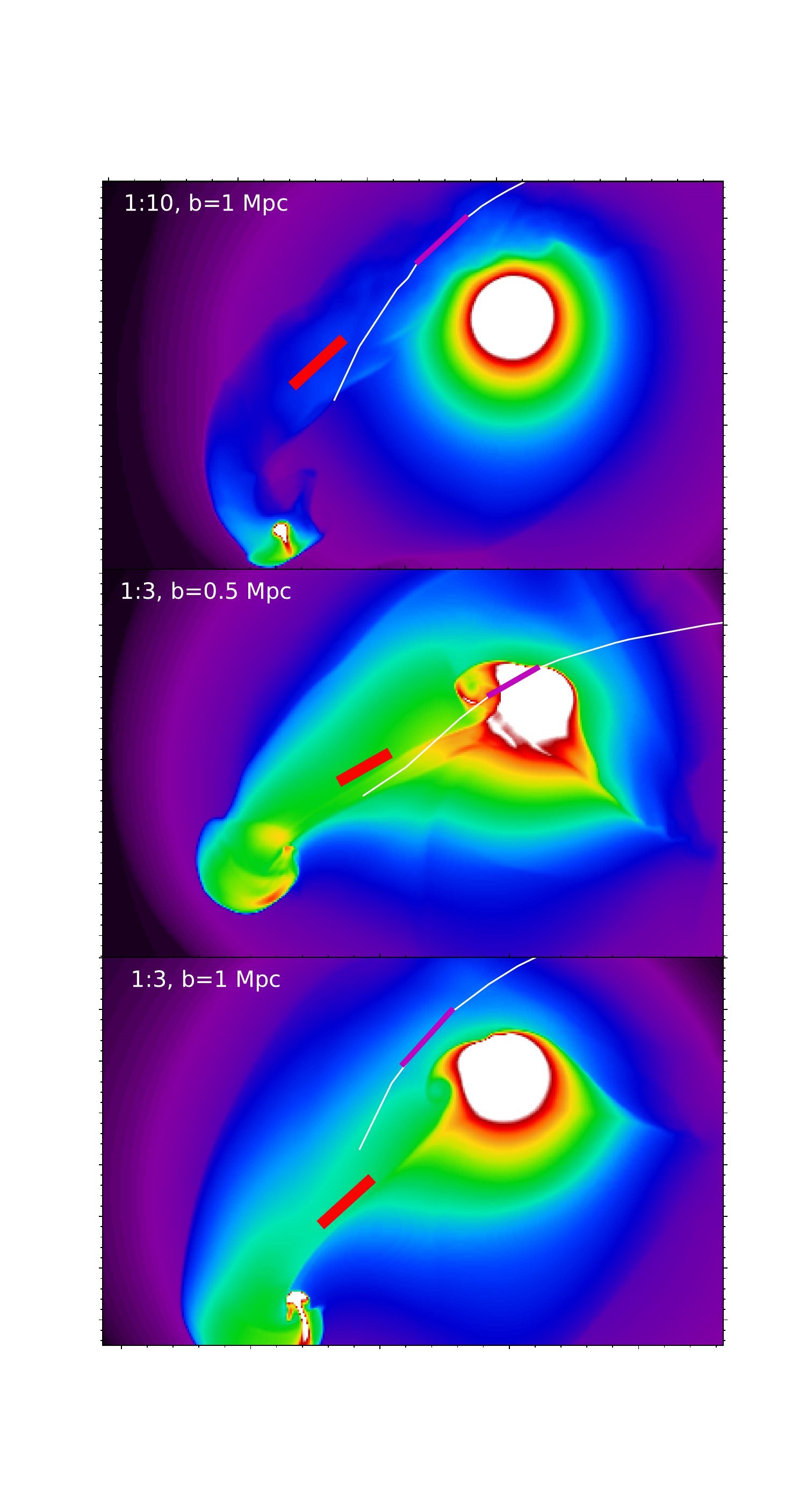}
    \flushleft
    \caption{The X-ray surface brightness images derived from numerical simulations \citep{2011ApJ...728...54Z,2018ApJS..234....4Z} illustrates the cluster merger occurring within the $x$-$y$ plane. The viewing angle is orthogonal to this plane, with the image captured approximately 0.56 Gyr post-core passage. 
    The inscriptions in white, located at the upper right corners, indicate the parameters of the simulation.
    The depicted colors emphasize the X-ray filamentary structures. The white trajectories denote the partial path of the subcluster relative to the rest frame of the principal cluster. Magenta solid lines mark the tangential alignment at the pericenters, while the major axis of the filament, calculated through SVD orthogonal regression, is indicated by the prominent red lines. {Alttext: three vertically aligned figures.}}
    \label{fig:filament}
\end{figure}

The numerical simulations \citep{2011ApJ...728...54Z,2018ApJS..234....4Z} predicted the presence of X-ray filamentary structures induced by cluster mergers.
The X-ray filament is a remnant of ram pressure stripping combined with tidal-induced spin motion at pericenter. At pericenter, the subcluster attains its maximum velocity, experiencing the strongest ram pressure in the dense ICM of the main cluster. The subcluster gas is stripped off by this ram pressure while being influenced by the tidal spin motion. Once the stripped gas is released from this rotational influence, it moves inertially. Consequently, the major axis of the X-ray filament does not coincide with the merger trajectory but is instead aligned with the tangential direction around pericenter. Figure \ref{fig:filament} shows three X-ray snapshots of numerical simulations from \citet{2011ApJ...728...54Z} and \citet{2018ApJS..234....4Z}. The major axis of the X-ray filament was measured in a manner similar to our analysis. We first defined a rectangular region of $100\times50$ pixels and applied SVD orthogonal regression to determine the angle and central position. This was done by testing various combinations of minimum and maximum threshold values. The prominent red lines indicate the mean characteristic, which deviates from the subcluster’s trajectory but aligns with the tangential direction at pericenter. The direction of the filament slightly changes with a merger time, but the overall features do not change. 

The coupling between the orbital trajectories and the tidal rotation of the main and subclusters is clearly found (e.g., in the middle and bottom panels of Figure \ref{fig:filament}): the bending directions of the tadpole-shaped gas in the main cluster and the comma-shaped gas in the subcluster are consistent with the azimuthal rotation of their trajectories. This coupling can be explained by the fact that the tidal torque acts to align the spin with the orbital motion, because the tidal bulge is slightly offset from the line connecting the two clusters, producing a torque that drives the rotation toward the trajectory direction. Consequently, the angular momentum induced by the mutual tidal interaction is coherently transferred to both the gas and dark matter components, leading to a systematic alignment between the orbital and spin directions
Notably, the relationship between the X-ray morphologies and the trajectories is similar to our result shown in the top panel of Figure \ref{fig:trajectory}. However, the orientations of the filaments and the lines connecting the two halos do not necessarily coincide.

% ==============================
% Remove Appendix mode
% ==============================
\clearpage
\makeatletter
  \@in@appendixfalse
  \let\appendix\relax
  \let\appendixname\relax
  \let\appendixtitle\relax
  \def\@sectitle@prefix{}%
  \everypar{}%
\makeatother
% ==============================

\bibliographystyle{apj}
\bibliography{pasj,hscrefs}

\end{document}